\documentclass[epj]{svjour}
\usepackage{graphics}
\usepackage{color}
\begin{document}
\title{The Letter ``S'' (and the sQGP)}
\author{J.L. Nagle
}                     
%
%
\institute{University of Colorado, Boulder CO 80309-0390}
\date{Received: August 29, 2006}
%
\abstract{
Data from the Relativistic Heavy Ion Collider over the last five years has led
many to conclude that the medium created is not the expected quark gluon plasma (QGP), but rather 
a strongly coupled or strongly interacting quark gluon plasma (sQGP).  We explore the
meaning of this possible paradigm shift and the experimental and theoretical arguments
that are associated with it.  In this proceedings we detail only a small subset of the relevant issues
as discussed at the Hot Quarks 2006 workshop.
\PACS{25.75.-q, 25.75.Nq}
} 
\maketitle
\section{Introduction}

The goal of this presentation at the Hot Quarks 2006 workshop was to attempt to develop a consistent understanding of the 
term ``sQGP'' and the physics conclusions that result.  The first step in achieving such a
goal is to detail what the letter ``s'' actually stands for and what is means.  
Does the terminology change from quark gluon plasma (QGP) to sQGP alphabetically symbolize an
important paradigm shift in the understanding of high temperature nuclear matter?

First, we detail what various people and collaborations have stated that ``sQGP'' means.
M. Gyulassy explained:  
``The name 'sQGP' (for strongly interacting Quark Gluon Plasma) helps to distinguish that matter from ordinary hadronic resonance matter (as described for example by RQMD) and
also from the original 1975 asymptotically free QGP (which I dubbed wQGP) that is now theoretically defined
in terms of re-summed thermal QCD~\cite{newdirections}.''  
Gyulassy and McLerran~\cite{gyulassymclerran} have argued 
``Our criteria for the discovery of QGP are (1) Matter at energy densities so large that simple
degrees of freedom are quarks and gluons.  This energy density is that predicted from lattice gauge theory for
the existence of a QGP in thermal systems, and is about 2 $GeV/fm^3$, (2) The matter must be to a good approximation
thermalized, (3) The properties of the matter associated with the matter while it is hot and dense must follow
QCD computations based on hydrodynamics, lattice gauge theory results, and perturbative QCD for hard processes 
such as jets.  All of the above are satisfied from the published data at RHIC...  This leads us to conclude that the 
matter produced at RHIC is a strongly coupled QGP (sQGP) contrary to original expectations that were based on 
weakly coupled plasma estimates.''

\begin{figure*}
\begin{center}
\resizebox{0.6\textwidth}{!}{%
  \includegraphics{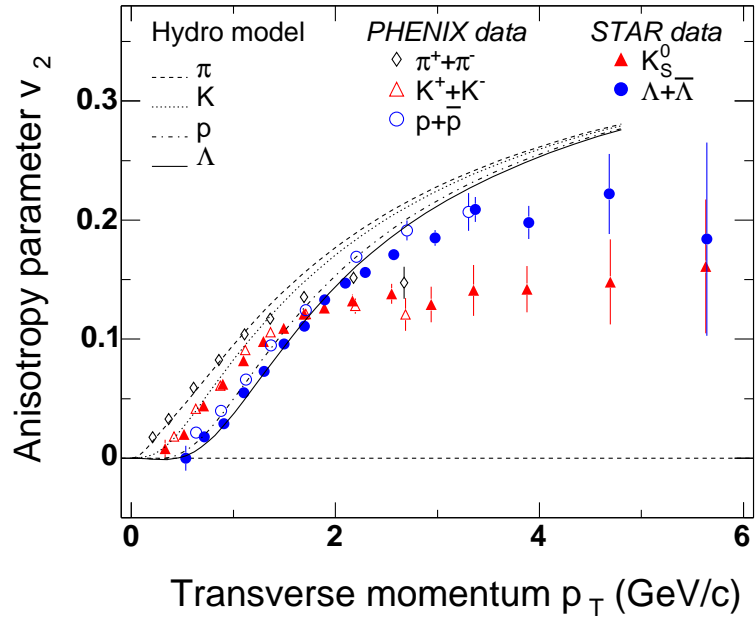}
}
\caption{Azimuthal anisotropy ($v_2$) as a function of $p_T$ from minimum bias gold-gold collisions.  Hydrodynamic calculations 
are shown as dashed lines.}
\label{fig:1}       
\end{center}
\end{figure*}

Although the estimates of the energy density at early times ($t=1~fm/c$) utilizing various methods disagree by
more than a factor of two~\cite{PHENIX_whitepaper}, 
all values are significantly above that predicted for the QGP phase transition for the first few $fm/c$.  For
example, the value from the Bjorken energy density equation is up to a factor of four lower than from hydrodynamic
calculations, but the Bjorken value is often viewed as a lower limit since it ignores any effects from 
longitudinal work.  Thus, the first criteria seems to be met.  Agreement of hydrodynamic calculations and
experimental data on transverse momentum spectra and in particular elliptic flow $v_2$ 
(see Figure~\ref{fig:1}~\cite{PHENIX_whitepaper,STARflow}) 
indicate very rapid equilibration times of order $t \approx 1~fm/c$~\cite{heinz}.  There have been questions
raised about the required degree of thermalization~\cite{borghini}; and, 
the originally stated agreement of hydrodynamics with the lattice equation of state (EOS) appears to
be overstated so that no quantitative constraint on latent heat or softness is yet warranted~\cite{PHENIX_whitepaper,pasi}.
However, it does appear that equilibration is approached more substantially than one might have expected 
from perturbative calculations (see later discussion
on this point).  Thus the first two criteria listed in ~\cite{gyulassymclerran} appear satisfied and 
might allow one to scientifically conclude that RHIC collisions have
created the QGP.  However, it is the critical third point that defines the experimental discovery of such.  

\section{Strongly interacting versus strongly coupled}

In the literature there is a mixture of terminology from strongly interacting and strongly coupled.  If it is strongly coupled, 
which coupling is being referred to?  In many talks and publications, the ``strongly coupled'' refers to the 
plasma coupling parameter $\Gamma$ (often used in the case of electromagnetic EM plasmas).  

\subsection{Plasma Coupling $\Gamma$}

\begin{figure*}
\begin{center}
\resizebox{0.6\textwidth}{!}{%
  \includegraphics{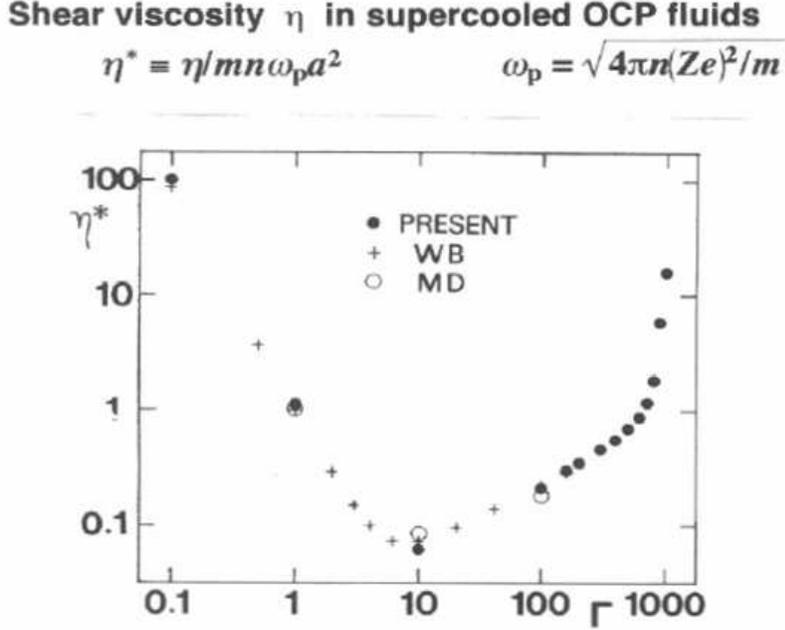}
}
\caption{Plotted is the scaled shear viscosity ($\eta^{*} = \eta/mn\omega_{p}a^{2}$) as a function of $\Gamma$ for
supercooled OCP fluids.}
\label{fig:2}       
\end{center}
\end{figure*}

This couping is defined as $\Gamma = <PE> / <KE>$, where PE is the average potential 
energy and KE is the average kinetic energy.  This parameter is used as a measure of the interaction strength in
EM plasmas.  Most EM plasmas that people are familiar with are weakly coupled plasmas where $\Gamma << 1$.  These 
behave like gases.  However, for $\Gamma >> 1$ the EM plasmas are strongly coupled and behave as low viscosity liquids and 
as solids at even larger $\Gamma$, as shown in Figure~\ref{fig:2}~\cite{ichimaru}.

Since EM plasmas have been widely studied, it is natural to seek to categorize the quark gluon plasma (QGP) in a similar fashion.
Recently at RHIC, there has been significant
publication on the QGP as a ``near-perfect liquid.''  Thus a question from someone outside the field of heavy ions is whether the
matter is in the plasma phase or liquid phase (often thought to be different regimes in the EM matter case).  
One must be careful about two different definitions of liquid being used here.  
Liquid can refer to a specific phase of electromagnetic matter and secondly where 
liquid refers to any matter whose dynamic evolution can be described by hydrodynamic equations of motion.
An EM plasma in the strong coupling
(large $\Gamma$ regime) is a plasma in that the electric charges are not confined to atoms, but has the liquid like property (second definition) of 
low viscosity.  
At RHIC, the matter produced shows some evidence of low viscosity (though not quantitative yet in terms of an upper limit on the shear viscosity).  
Thus, it may be a liquid (by the second definition), but may not share other EM liquid phase (first definition) properties.  
For example, many electromagnetic liquids are also highly incompressible.  For
the QGP, at baryon chemical potential $\mu_{B} = 0$ the pressure (P) and volume (V) are independent.  Again, the matter shares a property, but
not all.

These analogies are often useful, but only if they lead to new insights, rather than just new declarations and new terminology.
One has to be careful to define which properties are analogous.  For example, QCD always has screening of long range color magnetic
fields which means even a weakly interacting (asymptotically free) QGP will be quite different from a weakly coupled EM plasma.  Also,
on short distance scales, color electric and magnetic fields can be of equal order.  

Some in the field have argued the following logic:  Since the matter produced at RHIC has a large $\Gamma$ value, it must be a plasma
(as a phase).  This leads to the very strong conclusion that the matter at RHIC is a plasma (meaning a deconfined plasma of quarks
and gluons).  However, though EM plasmas are categorized in terms of $\Gamma$, not all large $\Gamma$ (i.e. low viscosity) matter
is a plasma at all.
As an example, there have been recent experiments with Lithium atoms where the mean free paths approach
zero under certain conditions~\cite{lithium}.  The Feshbach resonance in binary collisions of these alkali atoms at ultra-cold
temperatures allow experimentalists to tune the interaction strength.   The measurements reveal low viscosity and
``flow'' reminiscent of that seen in RHIC collisions.  However, these atoms are clearly not an EM plasma.  
Thus, at RHIC, demonstrating low viscosity does not 
prove the matter is a plasma.

One can push the plasma analogy and attempt to estimate the value of the $\Gamma$ parameter for the QGP and then 
attempt to infer other properties of the medium.  One such estimate~\cite{thoma} yields:
\begin{equation}
\Gamma = {{<PE>} \over {<KE>}} \approx {{\alpha_{s}/r} \over {3T}} \approx {{\alpha_{s}T} \over {3T}} \approx \alpha_{s}
\end{equation}
then utilizing the relation $\alpha_s = g^{2}(T)/4\pi$ and putting back in $d$ the characteristic inter-particle distance, one obtains:
\begin{equation}
\Gamma = {{Cg^{2}} \over {4 \pi d T}} \approx 1.5-5
\end{equation}
Note that this result is different from an earlier much larger estimate which had a factor of $4\pi$ unit error and was without
a factor of two scale-up for the approximately equal strength color magnetic interaction~\cite{thoma}.  Thoma notes that for EM plasmas ``the 
phase transition to the gas phase, assumed to happen at $\Gamma_c \approx 1$, takes place now at a few times the
transition temperature [from the QGP liquid to the QGP gas~\cite{thoma}.''  Note the title of this
article is ``The Quark-Gluon Plasma Liquid.''  
In the PHENIX whitepaper it states
``considerations such as these have led some to denote QGP in this regime as 'sQGP' for strongly interacting QGP~\cite{PHENIX_whitepaper}.''

In a recent set of papers~\cite{shuryak_cqgp}, the authors invoke a model referred to as cQGP where they calculate the shear viscosity as a function
of the dimensionless $\Gamma$ parameter.  The calculation seems to show a QGP with liquid like behavior (low viscosity) at large $\Gamma$
and an indication of solid behavior at even larger $\Gamma$, as was seen in the EM plasma case.  There has been speculation that
the QGP formed in heavy ion collisions could have crystalline or polymer chain type solid structures~\cite{shuryak_qm05}.  However, it
is critical to note that the letter 'c' stands for classical.  Thus, the entire calculation is done in the non-relativistic, non-quantum
regime and thus the possible insights gained have to be viewed with skepticism.  

The entire utilization of $\Gamma$ raises some significant questions.  The potential energy is taken as the Coulomb (short range) part
of the QCD potential as $\alpha_{s}/r$.  Unfortunately, when one has a system of (nearly) massless, relativistic particles then the
potential energy is not a well defined concept in a relativistic Quantum Field Theory (QFT).  This issue applies to a QFT for QED or QCD, but
is of particular concern for the QGP case here since anywhere near the transition temperature the light quarks are relativistic.  
The fundamental problem is that there is no unique distinction between the 
particles and the fields and thus no unique manner of separating potential energy and kinetic energy.  In which category do the
gluons belong for example?  In the case of heavy quarks, one might approximate them as static source charges and thus have a reasonable
attempt at separating the potential energy.  However, this is not the case for the QGP overall, and the assumption of a non-relativistic
limit in the cQGP case discussion above is not close to the real case for the QGP even near the critical temperature T = 170~MeV.
There are attempts to formulate an alternative for calculating $\Gamma$~\cite{jacak}.

Many people are interested in the $\Gamma$ calculation since it is how many EM plasmas are categorized.  However, other perfectly well-defined
in hydrodynamics and in a QFT measures of the interaction strength do exist that can alternatively be used.

\subsection{Shear Viscosity over Entropy Density $\eta/s$}

\begin{figure*}
\begin{center}
\resizebox{0.6\textwidth}{!}{%
\includegraphics{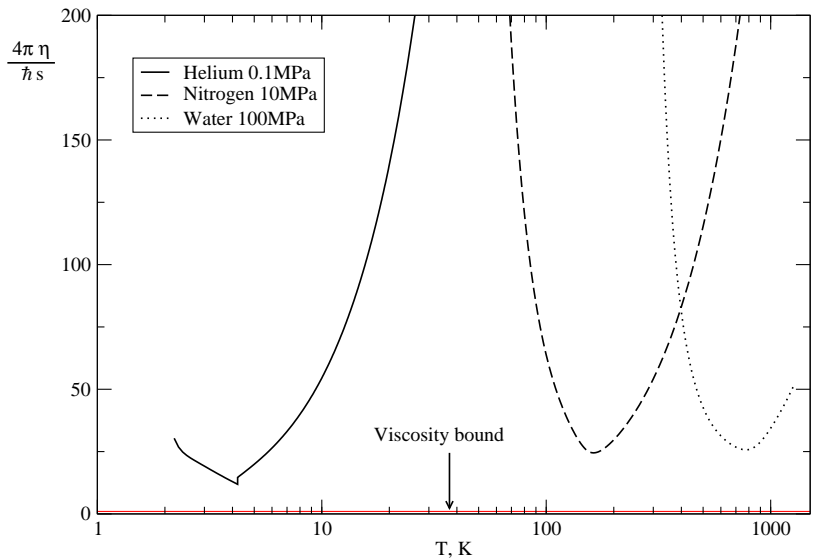}
}
\caption{Plotted are the shear viscosity to entropy density ratios ($\eta/s$) divided by
the conjectured lower bound as a function of temperature in Kelvin.  Shown are curves for
helium, nitrogen and water.}
\label{fig:3}       
\end{center}
\end{figure*}

There is a well defined measure of the interaction strength.  It is the ratio of the shear viscosity (a measure of the
mean free path of particles) and its entropy density (measure of the inter particle distances).  It is in fact this ratio 
$\eta / s$ that may be very small in the QGP as inferred from  hydrodynamic calculations and their comparison to experimental data.
Recent measurements of charm quark suppression at moderate $p_T ~\approx 2-5~GeV/c$ and non-zero elliptic flow $v_{2}$, may give
the best constraint on the diffusion coefficient from heavy quarks and subsequently $\eta/s$~\cite{mooreteaney,naglevienna}.  Full three-dimensional viscous
hydrodynamic calculations in comparison with precision data are needed to set a quantitatively reliable limit on $\eta/s$.  
Lattice simulations are presently unable to make reliable predictions of most dynamical properties of the quark-gluon
plasma. The calculation of phenomenologically relevant transport properties, 
such as the shear viscosity or collective modes, remains an important 
challenge \cite{Petreczky:2005zy}.

However, recently there has been important progress in calculating 
these dynamical properties perturbatively in a dual quantum field theory 
involving black holes in anti-de Sitter (AdS) space \cite{Kovtun:2004de}.  
This approach is based on the insight derived from string theory that 
weakly coupled gravity theories in higher dimensions can be dual to 
four-dimensional gauge theories in the strong coupling limit \cite{Maldacena:1997re}.  It must
be emphasized that these AdS/CFT (conformal field theory) techniques 
presently have the limitation that no higher dimensional gravity or 
string theory is known which is dual to QCD.  Work by Son {\it et al.} indicate
that there may be a lower viscosity bound $\eta/s > 1/4\pi$ applicable
for all systems including the quark gluon plasma.  A critical goal for
the field is to put the QCD matter data point on a plot like the one shown
in Figure~\ref{fig:3} for other systems~\cite{Kovtun:2004de}.  

An interesting side note is that in the figure these systems have a minimum
in the ratio $\eta/s$.  In fact, for helium, super-fluidity sets in at approximately
2 Kelvin, which is below the minimum.  The minimum occurs around 4 Kelvin which
is the gas to liquid phase transition point.  Thus the minimum is not a minimum
in viscosity, but rather the sudden change in entropy associated with the phase
transition.  Note the recent paper on the subject~\cite{larry}.

The most common example of a very low viscosity (or near perfect) fluid are the cases
shown in Figure~\ref{fig:3} which are referred to as super-fluids.  In most cases this
super-fluidity comes about from quantum mechanical effects dealing with the limited 
excitations at low temperature.  This seems quite different from the system at RHIC and
thus though there are many examples in the literature describing the matter at RHIC as a 
near perfect fluid, it is not termed a super-fluid.

\subsection{Strong Coupling $\alpha_s$}

\begin{figure*}
\begin{center}
\resizebox{0.6\textwidth}{!}{%
  \includegraphics{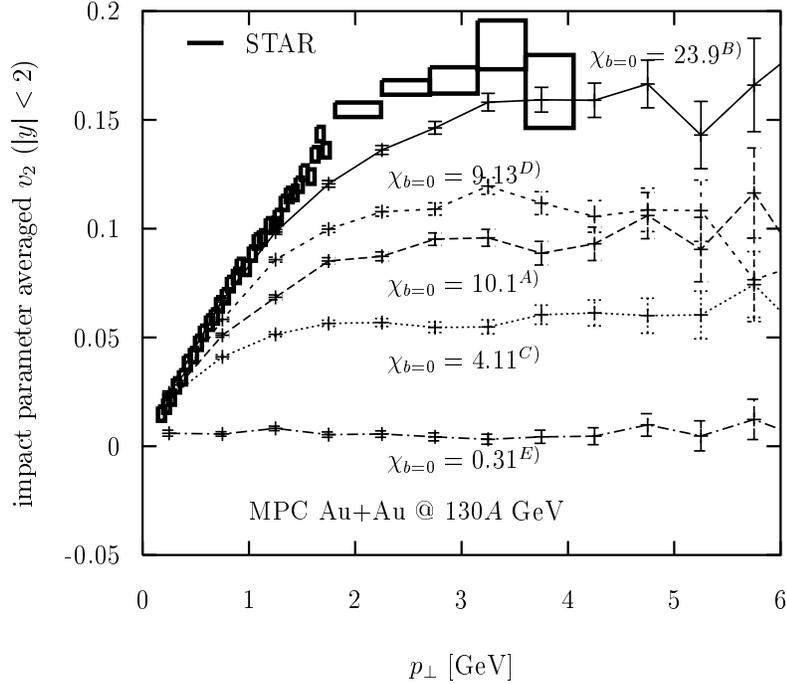}
}
\caption{Impact parameter averaged gluon elliptic flow as a function
of $p_T$ for Au+Au reactions at $\sqrt{s_{NN}}=130~GeV$ from MPC with various
values of the transport opacity for b=0.  Also shown are data points
from the STAR experiment.}
\label{fig:4}       
\end{center}
\end{figure*}

Another interpretation of the letter ``s'' is strongly coupled in the sense of
a large QCD coupling $\alpha_s$.  Clearly $\alpha_s$ is always, in any experimentally
accessible energy range, much greater than $\alpha_{EM} = 1/137$.  The wQGP, where 
the letter ``w'' stands for weak coupling, implies that perturbative expansions should
converge as $\alpha_s << 1$.  By contrast, sQGP would simply imply that perturbative 
techniques would not be applicable.  U. Heinz observed that 
``perturbative mechanisms seem unable to explain the phenomenological required
very short thermalization time scale, pointing to strong non-perturbative dynamics
in the QGP even at or above $2 \times T_c$.''~\cite{uli}.

In specific, analytic calculations utilizing perturbative expansions of
gluon scattering lead to long equilibration times ($ > 2.6 fm/c$) and thus rather modest
elliptic flow (i.e. small $v_2$)~\cite{baier}.  There are also numerical simulations that give similar 
results utilizing a $2 \rightarrow 2$ cross section of approximately 3 mb, as shown in Figure~\ref{fig:4}~\cite{molnar}.
One can artificially increase the cross section (or transport opacity) to match the data and it requires an order of
magnitude increase in the cross section.  In this sense, it is not a wQGP.  There are two important
caveats on these calculations.  One is that the Equation of State is too hard relative to lattice
results for the QGP.  More importantly is that there is some controversy over the inclusion of $2 \rightarrow 3$ and $3 \rightarrow 2$
processes.  Z.Xu {\it et al}~\cite{zhu} claim that their inclusion results in a dramatic decrease in the equilibration time
and thus a large increase in $v_2$.  At this conference it became clear that the critical part of
their result is that in $2 \rightarrow 3$ that the resulting gluons are emitted isotropically.  Under
this assumption it is easy to see why it leads to rapid isotropization.  Other implementations of these
processes show much smaller effects, in large part due to forward peaking of the emission
distribution.  This issue needs to be resolved.

In the third category used by Gyulassy and McLerran for discovery of the QGP, they cite utilizing 
perturbative methods to understand jet probes.
Radiative energy loss calculations are done perturbatively to describe the jet quenching phenomena.  In
fact, the calculations are effectively leading order.  GLV~\cite{glv}, for example, assumes the correct pQCD interaction
strength (noting that some calculations use a fixed couping $\alpha_s$ and others running), and then determine the color charge
density.  One obtains a result for $dN/dy({\rm gluons}) = 1000$ or $dN/dy({\rm quarks,gluons}) = 2000$.  The final entropy density
dS/dy is of order 5000, and thus since the entropy cannot be larger at earlier times it translates roughly into a 
limit $dN/dy({\rm quarks,gluons}) < 1300$~\cite{muller_annualreview}.  
One possibility is that more than just radiative energy loss contributes as has been highlighted by recent heavy quark results (perhaps indicating collisional energy loss).  However, another
approach is to say you know the color charge density and can then infer the coupling strength.  
This then implies that the coupling strength is much larger than predicted from the effectively leading
order perturbative calculation - which may be consistent with the sQGP description.  

\subsection{Bound States}

This strong coupling $\alpha_s$ is taken by Shuryak and collaborators~\cite{shuryak_bound} to imply that
the interaction between quasi-particles is strong enough to bind them.  Thus the sQGP
is composed of bound (not necessarily color neutral) $qq$, $q\overline{q}$, $gg$, $qg$, 
etc. states.  
However, recent lattice calculations for Baryon number - Electric charge correlations show no
such quasi-particles with these quantum numbers~\cite{karsch}.  It appears that lattice QCD is
ruling out $qq$ and $q\overline{q}$ states, though the results can say nothing about states without
these quantum numbers like $qg$ and $gg$ states.  

\subsection{Expectations}

A reasonable question is why there was an original expectation for a wQGP or perturbative plasma.  
``For plasma conditions realistically obtainable in nuclear collisions ($T \approx 250~MeV$, g = $\sqrt{4\pi\alpha_s}$)
the effective gluon mass $mg^{*} \approx 300~$MeV.   We must conclude, therefore, that the notion of
almost free gluons (and quarks) in the high temperature phase of QCD is quite far from the truth.  Certainly 
one has $mg^{*} << T$ when $g<<1$, but this condition is never really satisfied in QCD, because
$g \approx 1/2$ even at the Planck scale ($10^{19}$~GeV).''~\cite{bmueller}.
Despite this observation, many noted that from lattice gauge theory results the value of
$\epsilon/T^{4}$ approaches 80\% of the non-interacting gas limit.  
Some viewed this as
indicating only weak interactions, while some in the 
lattice community already thought that this 20\% difference from the Stefan Boltzmann limit 
was the effect of strong residual interactions in a non-perturbative system.
Also, recent results from AdS/CFT have
shown that one can be at the 80\% limit and still be in the very strongly interacting limit.

\section{Summary}

Exciting results of emergent phenomena at RHIC such as strong flow and jet quenching have sparked a great deal
of very positive new thinking about the medium created in these collisions.  It appears to represent a paradigm shift, 
although the earlier paradigm of a perturbatively describable (asymptotically free) plasma seems to have been poorly 
motivated.
F. Karsch puts it best: ``I do not really care what the 's' in sQGP means.  However, I am worried and partly also disappointed about
the way this new name is used.  The disappointment, of course, arises from the fact that suddenly a new name
seems to be necessary to describe the properties of QCD in a temperature regime which lattice gauge theory since
a long time have identified as 'not being an ideal gas' and 'impossible to be described by perturbation theory~\cite{newdirections}.''

As the field of heavy ions progresses, a coherent picture of the medium created may be emerging.  At this point there
are many ideas, some commensurate and other incommensurate with each other.  
Hopefully the future 
will tell us which are correct.


\section{Acknowledgment}

We thank the workshop
organizers for providing an environment for stimulating discussion and new ideas from young people.  We also acknowledge useful discussions prior to this workshop at the Boulder Workshop 2 and useful comments by one anonymous referee.  We acknowledge support from the United States Department of Energy grant DE-FG02-00ER41152.  
%
%

\end{document}